\documentclass[11pt]{amsart}
\usepackage{amssymb}

\usepackage{graphicx}
\usepackage{amscd}

\newtheorem{theorem}{Theorem}
\theoremstyle{plain}

\newtheorem{corollary}{Corollary}

\newtheorem{definition}{Definition}

\newtheorem{proposition}{Proposition}

\numberwithin{equation}{section}

\input{tcilatex}

\begin{document}
\title[The underlying digraphs of a coined quantum random walk]{The underlying digraphs \\
of a coined quantum random walk}
\author{Simone Severini}
\address{Department of Computer Science, University of Bristol, Merchant Venturers'
Building, Woodland road, BS8 1UB, Bristol, United Kingdom}
\email{severini@cs.bris.ac.uk}
\date{September 2002}

\begin{abstract}
We give a characterization of the line digraph of a regular digraph. We make
use of the characterization, to show that the underlying digraph of a coined
quantum random walk is a line digraph. We remark the connection between line
digraphs and in-split graphs in symbolic dynamics. (MSC2000: 05C50, 81P68)
\end{abstract}

\maketitle

\section{Introduction}

In this paper, we give a characterization of the line digraph of a regular
digraph. We make use of the characterization, to show that the underlying
digraph of a coined quantum random walk is a line digraph.

The structure of the paper is the following. In Section 2, we recall the
definition of line digraph and survey some properties. In Section 3, we give
the characterization. In Section 4, we consider coined quantum random walks
and remark the connection between these objects and line digraphs. Finally,
we remark the connection between line digraphs and in-split graphs as
defined in symbolic dynamics. The reader familiar with the general
properties of line digraphs can skip Section 2.

\section{Line digraphs}

\subsection{Definition}

The notion of line digraph has been introduced by Harary and Nornam in 1960 
\cite{HN60}. A classic survey on line graphs and digraphs is \cite{HB78}; a
recent one is \cite{P95}. Line digraphs are used in the design and analysis
of interconnection networks (see \emph{e.g.} \cite{FYA84}) and as a tool in
algorithms for DIRECTED\ MAX-CUT \cite{CE90}, certain poset problems \cite
{Sy84}, and special cases of TRAVELLING\ SALESMAN \cite{GKWS98}.

A \emph{directed graph}, for short \emph{digraph}, consists of a non-empty
finite set of elements called \emph{vertices} and a finite set of ordered
pairs of vertices called \emph{arcs}. Let us denote by $D=\left( V,A\right) $
a digraph with vertex-set $V\left( D\right) $ and arc-set $A\left( D\right) $%
. In an arc $\left( v_{i},v_{j}\right) $, $v_{i}$ and $v_{j}$ are called 
\emph{end-vertices} of $\left( v_{i},v_{j}\right) $; $v_{i}$ \emph{tail} and 
$v_{j}$ \emph{head} of $\left( v_{i},v_{j}\right) $. A digraph $D$ is an 
\emph{empty-graph} if $A\left( D\right) $ is the empty-set.

\begin{definition}[Line digraph]
The \emph{line digraph} of a digraph $D$ is denoted by $\overrightarrow{L}D$
and defined as follows: the vertex set of $\overrightarrow{L}D$ is $A\left(
D\right) $ and, for every $v_{h},v_{i},v_{j},v_{k}\in V\left( D\right) $, $%
\left( v_{h},v_{i}\right) ,\left( v_{j},v_{k}\right) \in A\left( 
\overrightarrow{L}D\right) $ if and only if $v_{i}=v_{j}$. The $k$\emph{%
-iterated line digraph} is recursively defined by 
\begin{equation*}
\overrightarrow{L}^{k}D=\overrightarrow{L}^{k-1}\overrightarrow{L}D.
\end{equation*}
\end{definition}

Now we need some standard terminology. 

A \emph{dipath} is a non-empty digraph $D$, where 
\begin{equation*}
\begin{tabular}{lll}
$V\left( D\right) =\left\{ v_{0},v_{1},...,v_{k}\right\} ,$ &  & $A\left(
D\right) =\left\{ \left( v_{0},v_{1}\right) ,\left( v_{1},v_{2}\right)
,...,\left( v_{k-1},v_{k}\right) \right\} ,$%
\end{tabular}
\end{equation*}
and, for every $v_{i},v_{j}\in V\left( D\right) $, $v_{i}\neq v_{j}$. A
dipath is called \emph{dicycle} if $v_{0}=v_{k}$. A $k$-\emph{dipath} ($k$-%
\emph{dicycle}), denoted by $\overrightarrow{P}_{n}$ ($\overrightarrow{C}_{n}
$), is a dipath (dicycle) on $n$ vertices. The undirected analogues of a
dipath and a dicycle are called \emph{path} and \emph{cycle}, respectively. 

A digraph $H$ is a s\emph{ubdigraph} of a digraph $D$ if $V\left( H\right)
\subseteq V\left( D\right) $, $A\left( H\right) \subseteq A\left( D\right) $
and every arc in $A\left( H\right) $ has both end-vertices in $V\left(
H\right) $. If $V\left( H\right) =V\left( D\right) $, $H$ is said to be a 
\emph{spanning subdigraph} of $D$. If every arc of $A\left( D\right) $ with
both end-vertices in $V\left( H\right) $ is in $A\left( H\right) $, we say
that $H$ is \emph{induced} by the set $X=V\left( H\right) $ and $H$ is an 
\emph{induced} subdigraph of $D$.

This is an obvious but important remark: the set of vertices $V\left( 
\overrightarrow{L}^{k}D\right) $ can be seen as the set of the $k$-dipaths
in $D$.

\subsection{General properties}

The theorems stated here are well-known; the proofs can be found in \cite
{HB78}.

For every $S\subset V\left( D\right) $, let 
\begin{equation*}
N_{D}^{-}\left( S\right) =\left\{ v_{i}:\left( v_{i},v_{j}\right) \in
A\left( D\right) ,v_{j}\in S\right\} 
\end{equation*}
and 
\begin{equation*}
N_{D}^{+}\left( S\right) =\left\{ v_{j}:\left( v_{i},v_{j}\right) \in
A\left( D\right) ,v_{i}\in S\right\} 
\end{equation*}
be the \emph{in-neighbourhood} and the \emph{out-neighbourhood} of $S$,
respectively. If the context is not equivoque we then write $N^{-}\left(
S\right) $ ($N^{+}\left( S\right) $). The \emph{in-degree} of $S$ is $%
d^{-}\left( S\right) =\left| N^{-}\left( S\right) \right| $; the \emph{%
out-degree} $d^{+}\left( S\right) =\left| N^{+}\left( S\right) \right| $.

A vertex $v_{i}$ in a digraph $D$ is \emph{isolated} if it is not in a
dipath with another vertex of $D$.

A digraph $D$ is \emph{connected} if, for every $v_{i},v_{j}\in V\left(
D\right) $, there is a dipath containing $v_{i}$ to $v_{j}$, or \emph{%
viceversa}; $D$ is \emph{strongly-connected} if, for every $v_{i},v_{j}\in
V\left( D\right) $, there is a dipath from $v_{i}$ to $v_{j}$ and from $%
v_{j} $ to $v_{i}$.

\begin{theorem}
Let $D$ be a digraph on $n$ vertices (none of which isolated)\ and $m$ arcs.
Then:

\begin{itemize}
\item[(i)]  
\begin{equation*}
\begin{tabular}{lll}
$\left| V\left( \overrightarrow{L}D\right) \right| =m$ & and & $\left|
A\left( \overrightarrow{L}D\right) \right| =\sum_{v_{i}\in V\left( D\right)
}d^{+}\left( v_{i}\right) d^{-}\left( v_{i}\right) ;$%
\end{tabular}
\end{equation*}

\item[(ii)]  
\begin{equation*}
\begin{tabular}{lll}
$N^{-}\left( \left( v_{i},v_{j}\right) \right) =d^{-}\left( v_{i}\right) $ & 
and & $N^{+}\left( \left( v_{i},v_{j}\right) \right) =d^{+}\left(
v_{j}\right) ;$%
\end{tabular}
\end{equation*}

\item[(iii)]  $\overrightarrow{L}D\cong \overrightarrow{P}_{n-1}$ if and
only if $D\cong \overrightarrow{P}_{n}$;

\item[(iv)]  $\overrightarrow{L}D\cong \overrightarrow{C}_{n}$ if and only
if $D\cong \overrightarrow{C}_{n}$.
\end{itemize}
\end{theorem}

\begin{theorem}
Let $D$ be a digraph. Then

\begin{itemize}
\item[(i)]  $\overrightarrow{L}^{n}D$ is an emty-graph, for some $n$, if and
only if $D$ has no dipaths;

\item[(ii)]  if $D$ has two dicycles joined by a $k$-dipath (possibly $k=1$%
), then 
\begin{equation*}
\lim_{n\rightarrow \infty }p_{n}=\infty ,
\end{equation*}
where $p_{n}$ is the number of vertices of $\overrightarrow{L}^{n}D$;

\item[(iii)]  if $D$ is strongly connected, and if $\overrightarrow{L}%
^{n}D\cong D$ for some $n$, then $\overrightarrow{L}D\cong D$, and $D$ is a
dicycle.
\end{itemize}
\end{theorem}

A digraph $D$ is \emph{hamiltonian}\ when $V\left( D\right) =V\left(
H\right) $, where $H$ is a dicycle. A digraph $D$ is \emph{eulerian} if it
is connected and, for every $v_{i}\in V\left( D\right) $, $d^{-}\left(
v_{i}\right) =d^{+}\left( v_{i}\right) $. A digraph $D$ is regular if, for
every $v_{i},v_{j}\in V\left( D\right) $, $d^{-}\left( v_{i}\right)
=d^{+}\left( v_{i}\right) =d^{-}\left( v_{j}\right) =d^{+}\left(
v_{j}\right) $.

\begin{theorem}
Let $D$ be a digraph. Then

\begin{itemize}
\item[(i)]  $\overrightarrow{L}D$ is strongly connected if and only if $D$
is strongly connected;

\item[(ii)]  $\overrightarrow{L}D$ is eulerian if and only if, for every arc 
$\left( v_{i},v_{j}\right) \in A\left( D\right) $, $d^{-}\left( v_{i}\right)
=d^{+}\left( v_{j}\right) $;

\item[(iii)]  $\overrightarrow{L}D$ is hamiltonian if and only if $D$ is
eulerian.
\end{itemize}
\end{theorem}

A \emph{general partition} of a set $S$ is a collection $\left\{
S_{i}\right\} _{i\in I}$ of (possibly empty) subsets of $S$, such that 
\begin{equation*}
\begin{tabular}{lll}
$S=\bigcup_{i\in I}S_{i},$ & and & $S_{i}\cap S_{j}=\emptyset $ if $i\neq j.$%
\end{tabular}
\end{equation*}

A digraph $D$ is said to be $F$\emph{-free} if it does not contain any
subdigraph isomorphic to $F$.

The \emph{adjacency matrix} of a digraph $D$ on $n$ vertices, denoted by $%
M\left( D\right) $ is the $n\times n$ $\left( 0,1\right) $-matrix with $ij$%
-th entry equal to $1$ if $\left( i,j\right) \in A\left( D\right) $, and
equal to $0$, otherwise. Let $r_{i}\left( M\right) $ and $c_{j}\left(
M\right) $ be respectively the $i$-th row and the $j$-th column of a matrix $%
M$. Let $\left\langle a,b\right\rangle $ the inner product of vectors $a$
and $b$.

\begin{theorem}
Let $D$ be a digraph. Then the following statements are equivalent:

\begin{itemize}
\item[(i)]  $D$ is a line digraph;

\item[(ii)]  there exist two general partitions $\left\{ A_{i}\right\}
_{i\in I}$ and $\left\{ B_{i}\right\} _{i\in I}$ of $V\left( D\right) $ such
that, for each $i$ and $j$, 
\begin{equation*}
\begin{tabular}{lll}
$\left| A_{j}\cap B_{i}\right| \leq 1-\delta _{i,j},$ & and such that & $%
A\left( D\right) =\bigcup_{i\in I}A_{i}\times B_{i};$%
\end{tabular}
\end{equation*}

\item[(iii)]  any two rows of $M\left( D\right) $ are identical or
orthogonal, $M_{i,i}=0$ for all $i$, and if $r_{i}\left( M\right)
=r_{j}\left( M\right) \neq 0$ then $\left\langle c_{i}\left( M\right)
,c_{j}\left( M\right) \right\rangle =0$;

\item[(iv)]  any two columns of $M\left( D\right) $ are identical or
orthogonal, $M_{i,i}=0$ for all $i$, and if $c_{i}\left( M\right)
=c_{j}\left( M\right) \neq 0$ then $\left\langle r_{i}\left( M\right)
,r_{j}\left( M\right) \right\rangle =0$;

\item[(v)]  $D$ is $D_{3}$- and $D_{4}$-free, where 
\begin{equation*}
D_{3}=\left( \left\{ 1,2,3,4\right\} ,\left\{ \left( 1,2\right) ,\left(
1,3\right) ,\left( 4,2\right) \right\} \right) ,
\end{equation*}
and 
\begin{equation*}
D_{4}=\left( \left\{ 1,2,3\right\} ,\left\{ \left( 1,2\right) ,\left(
3,1\right) ,\left( 3,2\right) \right\} \right) .
\end{equation*}
Any digraph obtained from $D_{3}$ or $D_{4}$ by adding arcs is not an
induced subdigraph of $D$, with the exception of 
\begin{equation*}
D_{3}^{\prime }=\left( \left\{ 1,2,3,4\right\} \left\{ \left( 1,2\right)
,\left( 1,3\right) ,\left( 4,2\right) ,\left( 4,3\right) \right\} \right) 
\end{equation*}
and 
\begin{equation*}
D_{4}^{\prime }=\left( \left\{ 1,2,3\right\} ,\left\{ \left( 1,2\right)
,\left( 3,1\right) ,\left( 3,2\right) ,\left( 1,1\right) \right\} \right) .
\end{equation*}
\end{itemize}
\end{theorem}

\begin{theorem}[\protect\cite{P01}]
Let $D$ be a strongly connected digraph. If $D$ is regular then all its
iterated line digraphs are regular, thus eulerian, and hamiltonian.
\end{theorem}

\subsection{Algebraic properties}

\subsubsection{Spectrum}

Let $D$ be a digraph on $n$ vertices. The set of eigenvalues of the
adjacency matrix $M\left( D\right) $ is denoted by $sp\left( D\right) $, and
called \emph{spectrum} of $D$ \cite{CDH95}. Let $I_{n}$ be the $n\times n$
identity matrix.

\begin{theorem}[see \protect\cite{LZ83} or \protect\cite{R01}]
\label{eig}Let $D$ be a digraph. The characterisitc polynomial of $%
\overrightarrow{L}D$ is 
\begin{equation}
P\left( \overrightarrow{L}D,x\right) =x^{\left| A\left( D\right) \right|
-\left| V\left( D\right) \right| }P\left( D,x\right) ,  \label{cp}
\end{equation}
where 
\begin{equation*}
P\left( D,x\right) =\det \left( xI_{n}-M\left( D\right) \right)
\end{equation*}
is the characteristic polynomial of $D$.
\end{theorem}

Let $\left\{ 0\right\} _{n}$ be a set of $n$ zeros. By Theorem \ref{eig},
since 
\begin{equation*}
sp\left( K_{d}^{+}\right) =\left\{ d\right\} \cup \left\{ 0\right\} _{d-1},
\end{equation*}
we have the following corollary.

\begin{corollary}
$sp\left( B\left( d,k\right) \right) =\left\{ d\right\} \cup \left\{
0\right\} _{d^{k}-1}$.
\end{corollary}

\subsubsection{Moore-Penrose inverse}

A matrix $M^{+}$ is a \emph{Moore-Penrose inverse} of a matrix $M$ if

\begin{itemize}
\item[(i)]  $MM^{+}M=M$,

\item[(ii)]  $M^{+}MM^{+}=M^{+}$,

\item[(iii)]  $\left( MM^{+}\right) ^{\intercal }=MM^{+}$,

\item[(iv)]  $\left( M^{+}M\right) ^{\intercal }=M^{+}M$.
\end{itemize}

\begin{theorem}[\protect\cite{Sm81}]
A square $\left( 0,1\right) $-matrix has a Moore-Penrose inverse if and only
if it is the adjacency matrix of a line digraph.
\end{theorem}

\subsubsection{Permanent}

The \emph{permanent} of a $\left( 0,1\right) $-matrix $M$ is 
\begin{equation*}
per\left( M\right) =\sum_{\pi \in S_{n}}\prod_{i=1}^{n}A_{i,\pi \left(
i\right) },
\end{equation*}
where $S_{n}$ is the symmetric group on an $n$-set.

\begin{theorem}[\protect\cite{KSW97}]
Let $D$ be a digraph. Then 
\begin{equation*}
per\left( M\left( \overrightarrow{L}D\right) \right) >0
\end{equation*}
if and only if each connected component of $D$ is eulerian.
\end{theorem}

\subsection{Line digraphs and unitary matrices}

Let $M$ be a matrix over any field. The \emph{support of }$M$ is the $\left(
0,1\right) $-matrix with $ij$-th element equal to $1$ if $M_{i,j}\neq 0$,
and equal to $0$, otherwise. The \emph{digraph of} $M$ is the digraph whose
adjacency matrix is the support of $M$. If a digraph $D$ is the digraph of a
matrix $M$ then we say that $D$, or indistinctly $M\left( D\right) $, \emph{%
supports} $M$.

\begin{theorem}[\protect\cite{Se03}]
Let $D$ be a digraph. Then $\overrightarrow{L}D$ is the digraph of a unitary
matrix if and only if every connected component of $D$ is eulerian.
\end{theorem}

\section{A characterization of the line digraph of a regular digraph}

In this section, we characterize the adjacency matrix of the line digraph of
a regular digraph. 

\subsection{The characterization}

Let $D$ be a digraph. A $1$\emph{-cycle factor} of $D$ is the disjoint union
of directed cycles spanning $D$. The adjacency matrix of a $1$-cycle factor
is a permutation matrix.

A $k$-\emph{factor} $F$ of $D$ is a $k$-regular spanning subdigraph of $D$.
A $k$-\emph{factorization of }$D$ is a set $\left\{
F_{1},F_{2},...,F_{m}\right\} $ of pairwise arc-disjoint $k$-factors of $D$ 
\emph{covering} $A\left( D\right) $, that is $A\left( D\right) =F_{1}\cup
F_{2}\cup ...\cup F_{m}$. More generally, a \emph{factorization of }$D$ is a
set of pairwise arc-disjoint factors of $D$, possibly of different degrees,
covering $A\left( D\right) $.

The \emph{growth} $\Upsilon _{D}\left( F\right) $ of $D$, introduced in \cite
{HS96}, is a digraph derived from a spanning subdigraph $F$ of $D$, by
adding, for each vertex $v_{i}\in F$, 
\begin{equation*}
\begin{tabular}{lll}
$l=\left| N_{D}^{+}\left( v_{i}\right) \right| -\left| N_{F}^{+}\left(
v_{i}\right) \right| $ & vertices & $\left\{ v_{1},...,v_{l}\right\} $%
\end{tabular}
\end{equation*}
and 
\begin{equation*}
\begin{tabular}{lll}
$l$ & arcs & $\left( v_{i},v_{1}\right) ,\left( v_{i},v_{2}\right)
,...,\left( v_{i},v_{l}\right) .$%
\end{tabular}
\end{equation*}

\begin{proposition}
\label{ma}Let $D$ be a $k$-regular digraph. Let $\left\{
F_{1},F_{2},...,F_{k}\right\} $ be a $1$-factorization of $D$. Then there is
a labeling of $\overrightarrow{L}D$, such that 
\begin{equation*}
M\left( \overrightarrow{L}D\right) =\left[ 
\begin{array}{cccc}
M\left( F_{1}\right) & M\left( F_{2}\right) & \cdots & M\left( F_{k}\right)
\\ 
M\left( F_{1}\right) & M\left( F_{2}\right) & \cdots & M\left( F_{k}\right)
\\ 
\vdots & \vdots & \ddots & \vdots \\ 
M\left( F_{1}\right) & M\left( F_{2}\right) & \cdots & M\left( F_{k}\right)
\end{array}
\right] .
\end{equation*}
\end{proposition}

\begin{proof}
Since $D$ is $k$-regular, the adjacency matrix of $D$ is 
\begin{equation*}
M\left( D\right) =\sum_{j=1}^{k}M\left( F_{j}\right) .
\end{equation*}
Label by the pair $\left( F_{j},v_{l}\right) $ the arc $\left(
v_{i},v_{l}\right) $ of $F_{j}$. Construct $\Upsilon _{D}\left( F_{j}\right) 
$. Note that $F_{j}$ is a subdigraph of $\Upsilon _{D}\left( F_{j}\right) $.
So, in $\Upsilon _{D}\left( F_{j}\right) $, we can keep the same labelling
of $F_{j}$. In addition, we label by $v_{m}$, with $m\neq l$, the $m$-th of
the $k-1$ vertices which are heads of the arcs incident to $v_{i}$ and that
are not in $F_{j}$. Then label the arc $\left( v_{i},v_{m}\right) $ with the
pair $\left( F_{m},v_{m}\right) $. Since each $F_{j}$ is the disjoint union
of dicycles, and since the line digraph of a dicycle is a dicycle (see, 
\emph{e.g.}, \cite{HB78}, Theorem 7.1), it follows that 
\begin{equation*}
\Upsilon _{D}\left( F_{j}\right) \cong \Upsilon _{\overrightarrow{L}D}\left(
F_{j}\right) .
\end{equation*}
Given the chosen labelling and an ordering of the vertices, 
\begin{equation*}
M\left( \Upsilon _{\overrightarrow{L}D}\left( F_{j}\right) \right) =\left[ 
\begin{array}{cccccc}
\mathbf{0} & \mathbf{0} & \cdots & \mathbf{0} & \cdots & \mathbf{0} \\ 
\vdots & \cdots & \cdots & \vdots & \ddots & \vdots \\ 
\mathbf{0} & \mathbf{0} & \cdots & \mathbf{0} & \cdots & \mathbf{0} \\ 
M\left( F_{1}\right) & M\left( F_{2}\right) & \cdots & M\left( F_{j}\right)
& \cdots & M\left( F_{k}\right) \\ 
\mathbf{0} & \mathbf{0} & \cdots & \mathbf{0} & \cdots & \mathbf{0} \\ 
\vdots & \cdots & \cdots & \vdots & \ddots & \vdots \\ 
\mathbf{0} & \mathbf{0} & \cdots & \mathbf{0} & \cdots & \mathbf{0}
\end{array}
\right] ,
\end{equation*}
where $M\left( F_{j}\right) $ is the $\left( j,j\right) $-th block of $%
M\left( \Upsilon _{\overrightarrow{L}D}\left( F_{j}\right) \right) $. Since,
the set 
\begin{equation*}
\left\{ F_{1},F_{2},...,F_{k}\right\}
\end{equation*}
is a $1$-factorization of $D$ if and only if 
\begin{equation*}
\left\{ \Upsilon _{\overrightarrow{L}D}\left( F_{1}\right) ,\Upsilon _{%
\overrightarrow{L}D}\left( F_{2}\right) ,...,\Upsilon _{\overrightarrow{L}%
D}\left( F_{k}\right) \right\}
\end{equation*}
is a $1$-factorization of $\overrightarrow{L}D$ \cite{HS96}, it follows that 
\begin{equation*}
M\left( \overrightarrow{L}D\right) =\sum_{j=1}^{k}M\left( \Upsilon _{%
\overrightarrow{L}D}\left( F_{j}\right) \right) .
\end{equation*}
\end{proof}

\subsection{Example}

Consider the digraph $D$ with adjacency matrix matrix 
\begin{equation*}
M\left( D\right) =\left[ 
\begin{array}{cccc}
0 & 1 & 1 & 0 \\ 
1 & 0 & 0 & 1 \\ 
1 & 0 & 0 & 1 \\ 
0 & 1 & 1 & 0
\end{array}
\right] .
\end{equation*}
Note that $D$ is a $2$-cube. Chose a $1$-factorization of $D$, $\left\{
F_{1},F_{2}\right\} $. For example, we chose $\left\{ F_{1},F_{2}\right\} $
such that 
\begin{equation*}
\begin{tabular}{lll}
$M\left( F_{1}\right) =\left[ 
\begin{array}{cccc}
0 & 1 & 0 & 0 \\ 
1 & 0 & 0 & 0 \\ 
0 & 0 & 0 & 1 \\ 
0 & 0 & 1 & 0
\end{array}
\right] $ & and & $M\left( F_{2}\right) =\left[ 
\begin{array}{cccc}
0 & 0 & 1 & 0 \\ 
0 & 0 & 0 & 1 \\ 
1 & 0 & 0 & 0 \\ 
0 & 1 & 0 & 0
\end{array}
\right] $%
\end{tabular}
\end{equation*}
We obtain 
\begin{equation*}
M\left( \Upsilon _{D}\left( F_{1}\right) \right) = 
\begin{array}{ccccccccc}
& F_{1},1 & F_{1},2 & F_{1},3 & F_{1},4 & F_{2},1 & F_{2},2 & F_{2},3 & 
F_{2},4 \\ 
F_{1},1 & 0 & 1 & 0 & 0 & 0 & 0 & 1 & 0 \\ 
F_{1},2 & 1 & 0 & 0 & 0 & 0 & 0 & 0 & 1 \\ 
F_{1},3 & 0 & 0 & 0 & 1 & 1 & 0 & 0 & 0 \\ 
F_{1},4 & 0 & 0 & 1 & 0 & 0 & 1 & 0 & 0 \\ 
F_{2},1 & 0 & 0 & \cdots &  &  &  & \cdots & 0 \\ 
F_{2},2 & 0 & 0 & \cdots &  &  &  & \cdots & 0 \\ 
F_{2},3 & \vdots & \vdots &  &  &  &  &  & \vdots \\ 
F_{2},4 & 0 & \cdots &  &  &  &  & \cdots & 0
\end{array}
\end{equation*}
and 
\begin{equation*}
M\left( \Upsilon _{D}\left( F_{2}\right) \right) = 
\begin{array}{ccccccccc}
& F_{1},1 & F_{1},2 & F_{1},3 & F_{1},4 & F_{2},1 & F_{2},2 & F_{2},3 & 
F_{2},4 \\ 
F_{1},1 & 0 & 0 & \cdots &  &  &  & \cdots & 0 \\ 
F_{1},2 & 0 & 0 & \cdots &  &  &  & \cdots & 0 \\ 
F_{1},3 & 0 & \vdots &  &  &  &  &  & \vdots \\ 
F_{1},4 & 0 & \cdots &  &  &  &  & \cdots & 0 \\ 
F_{2},1 & 0 & 1 & 0 & 0 & 0 & 0 & 1 & 0 \\ 
F_{2},2 & 1 & 0 & 0 & 0 & 0 & 0 & 0 & 1 \\ 
F_{2},3 & 0 & 0 & 0 & 1 & 1 & 0 & 0 & 0 \\ 
F_{2},4 & 0 & 0 & 1 & 0 & 0 & 1 & 0 & 0
\end{array}
.
\end{equation*}
Now, 
\begin{equation*}
\begin{tabular}{lll}
$\Upsilon _{D}\left( F_{1}\right) \cong \Upsilon _{\overrightarrow{L}%
D}\left( F_{1}\right) $ & and & $\Upsilon _{D}\left( F_{2}\right) \cong
\Upsilon _{\overrightarrow{L}D}\left( F_{2}\right) $%
\end{tabular}
\end{equation*}
Since $\left\{ \Upsilon _{\overrightarrow{L}D}\left( F_{1}\right) ,\Upsilon
_{\overrightarrow{L}D}\left( F_{2}\right) \right\} $ is a factorization of $%
\overrightarrow{L}D$, 
\begin{eqnarray*}
M\left( \overrightarrow{L}D\right) &=&M\left( \Upsilon _{D}\left(
F_{1}\right) \right) +M\left( \Upsilon _{D}\left( F_{2}\right) \right) = \\
&=&\left[ 
\begin{array}{cc}
M\left( F_{1}\right) & M\left( F_{2}\right) \\ 
M\left( F_{1}\right) & M\left( F_{2}\right)
\end{array}
\right] .
\end{eqnarray*}

\section{Line digraphs and coined quantum random walks}

In this section, we first recall the definition of coined quantum random
walk, then, we show that the underlying digraph of a coined quantum random
walk is a line digraph. Coined quantum random walks, and continuous-time
quantum random walks, are surveyed in \cite{K03}.

\subsection{Definition}

Let $D$ be a $k$-regular digraph on $n$ vertices. Consider two quantum
systems, to which are respectively assigned the Hilbert spaces $\mathcal{H}%
^{k}$ and $\mathcal{H}^{n}$, of respective dimensions $k$ and $n$. Label
each ray of the standard basis of $\mathcal{H}^{k}$ by a $1$-factor of $D$.
Label each ray of the standard basis of $\mathcal{H}^{n}$ by the vertices of 
$D$. Let 
\begin{equation*}
\left\{ \left| F_{j},v_{i}\right\rangle :0\leq j\leq k,0\leq i\leq n\right\}
\end{equation*}
be the standard basis of $\mathcal{H}^{k\cdot n}=\mathcal{H}^{k}\otimes 
\mathcal{H}^{n}$. Let $\left| \psi _{t}\right\rangle \in \mathcal{H}^{k\cdot
n}$ be the state of the system at time $t$. Let 
\begin{equation*}
\begin{tabular}{lll}
$\widehat{C}:\mathcal{H}^{k}\longrightarrow \mathcal{H}^{k}$ & and & $%
\widehat{T}:\mathcal{H}^{k\cdot n}\longrightarrow \mathcal{H}^{k\cdot n}$%
\end{tabular}
\end{equation*}
be unitary operators, such that 
\begin{equation*}
\begin{tabular}{lll}
$\widehat{C}:\left| F_{j}\right\rangle \longrightarrow \sum_{j=1}^{k}\alpha
_{j}\left| F_{j}\right\rangle $ & and & $\widehat{T}:\left|
F_{j},v_{i}\right\rangle \longrightarrow \left| F_{j},v_{l}\right\rangle $,
where $\left( v_{i},v_{l}\right) \in F_{j}$.
\end{tabular}
\end{equation*}
The matrices $C$ and $T$, arising from these operators, are respectively
called \emph{coin} and \emph{shift}. Define the unitary operator 
\begin{equation*}
\widehat{U}:\left| F_{j},v_{i}\right\rangle \longrightarrow
\sum_{j=1,l:\left( v_{i},v_{l}\right) \in F_{j}}^{k}\alpha _{j}\left|
F_{j},v_{l}\right\rangle .
\end{equation*}
A \emph{coined quantum random walk} on $D$\ \emph{with coin} $C$, induced by
the\ \emph{transition matrix} 
\begin{equation*}
U=T\cdot \left( C\otimes I_{n}\right) ,
\end{equation*}
is a sequence $\left\{ X_{t}\right\} $ of random variables. The sequence
starts at $X_{0}=v_{i}$, fixed, or drawn from some initial distribution. At
time $t$, the probability that $X_{t}=v_{j}$, conditioned on a initial state 
$\left| \psi _{0}\right\rangle $, is 
\begin{equation*}
\begin{tabular}{lll}
$\Pr\nolimits_{t}\left( v_{j}|\psi _{0}\right) =\sum_{i}\left\langle \psi
_{t}\right| P^{\left( i,j\right) }\left| \psi _{t}\right\rangle ,$ & where & 
$\left| \psi _{t}\right\rangle =U^{t}\left| \psi _{0}\right\rangle $%
\end{tabular}
\end{equation*}
and $P^{\left( i,j\right) }$ is the projector onto the $k$-dimensional
subspace spanned $\left| F_{j},v_{i}\right\rangle $.

\subsection{The underlying digraph of a coined quantum random walk}

The \emph{underlying digraph} of a random walk induced by a transition
matrix $M$ is the digraph of $M$. In this sense, the underlying digraph of a
coined quantum random walk induced by a unitary matrix $U$ is the digraph of 
$U$.

\begin{proposition}
The underlying digraph of a coined quantum random walk is a line digraph.
\end{proposition}

\begin{proof}
Let $D$ be a $k$-regular digraph on $n$ vertices. Let $\left\{
F_{1},F_{2},...,F_{k}\right\} $ be a factorization of $D$. By Proposition 1, 
\begin{equation*}
M\left( \overrightarrow{L}D\right) =\left( M\left( K_{k}^{+}\right) \otimes
I_{n}\right) \cdot T,
\end{equation*}
where 
\begin{equation*}
T=\bigoplus_{j=1}^{k}M\left( F_{j}\right) ,
\end{equation*}
Observe that 
\begin{equation*}
M\left( \overrightarrow{L}D\right) ^{\intercal }=T\cdot \left( M\left(
K_{k}^{+}\right) \otimes I_{n}\right) .
\end{equation*}
If in this equation we replace $M\left( K_{k}^{+}\right) $ with a unitary
matrix $C$ of size $k$, we obtain 
\begin{equation*}
T\cdot \left( C\otimes I_{n}\right) ,
\end{equation*}
which is the transition matrix of a coined quantum random walk on $D$.
\end{proof}

\subsection{Example:\ the cycle}

The following construction is described in \cite{FFY92}. 

Let $D=Cay\left( G,S\right) $ be a Cayley digraph and let $S=\left\{
s_{1},s_{2},...,s_{d}\right\} $. Let 
\begin{equation*}
H=\left\{ \pi _{s_{i}}:s_{i}\in S\right\} 
\end{equation*}
be a set of permutations of $S$. Let $H$ act on $S$ as a regular group: we
take $\pi _{s_{1}}$ as identity element and, for every $s_{i},s_{j}\in S$,
we assume that there exists a unique $s_{k}$, such that $\pi _{s_{k}}\left(
s_{i}\right) =s_{j}$. The set $H$ can be used to define the following
operation on $S$: 
\begin{equation*}
s_{i}\circledcirc s_{j}=\pi _{s_{i}}\left( s_{j}\right) .
\end{equation*}
Let $S^{\ast }\subseteq S$. The elements $s_{k}\in S^{\ast }$, such that,
for all $s_{i},s_{j}\in S$, 
\begin{equation*}
s_{k}\circledcirc \left( s_{i}\circledcirc s_{j}\right) =\left(
s_{k}\circledcirc s_{i}\right) \circledcirc s_{j},
\end{equation*}
form a group $\left( S^{\ast },\circledcirc \right) $, whose identity
element is $s_{1}$.

Let $\times _{sd}$ be the symbol of the (external) semidirect product. If $%
H\leq Aut\left( G\right) $ then $\left( S,\circledcirc \right) \cong H$. In
such a case, the digraph $\overrightarrow{L}D$ is a Cayley digraph of 
\begin{equation*}
\Omega \cong G\times _{sd}\left( S,\circledcirc \right) ,
\end{equation*}
with respect to the set of generators 
\begin{equation*}
\left\{ \left( s_{1},s_{i}\right) :s_{i}\in S\right\} .
\end{equation*}

Let $C_{n}$ be the cyclic group of order $n$. Take $n$ odd. Assume that $%
\mathbb{Z}_{n}\cong C_{n}$ is generated by the set $S=\left\{ 1,n-1\right\} $%
. The Cayley digraph $D=Cay\left( \mathbb{Z}_{n},S\right) $ is a cycle of
lenght $n$. Let $H=\left\{ \pi _{1},\pi _{n-1}\right\} $. Let $\pi _{1}$ be
the identity and, 
\begin{equation*}
\begin{tabular}{lll}
$\pi _{n-1}\left( n-1\right) =1$ & and & $\pi _{n-1}\left( 1\right) =n-1.$%
\end{tabular}
\end{equation*}
Then $H\cong \mathbb{Z}_{2}$. The element $\pi _{n-1}\in Aut\left( \mathbb{Z}%
_{n}\right) $. Then $\overrightarrow{L}D$ is the Cayley digraph of the group 
\begin{equation*}
\Omega \cong \mathbb{Z}_{n}\times _{sd}\mathbb{Z}_{2}\cong D_{n},
\end{equation*}
where 
\begin{equation*}
D_{n}=\left\langle a,b:a^{n}=b^{2}=e,a^{n-1}=bab\right\rangle 
\end{equation*}
is the dihedral group of order $2n$, generated by its standard presentation. 

By denoting a permutation in the standard cycle notation, we write 
\begin{equation*}
\begin{tabular}{lll}
$g=\left( 1\text{ }2\text{ }3\text{ ... }n\right) $ & and & $g^{n-1}=\left( 1%
\text{ }n\text{ }n-1\text{ ... }2\text{ }1\right) $,
\end{tabular}
\end{equation*}
where $g=1$ and $g^{n-1}=n-1$. Then 
\begin{equation*}
\begin{tabular}{lll}
$\pi _{n-1}\left( 1\right) =\sigma \left( 1\right) \sigma =n-1,$ & and & $%
\pi _{1}\left( n-1\right) =\sigma \left( n-1\right) \sigma =1,$%
\end{tabular}
\end{equation*}
where 
\begin{equation*}
\sigma =\left( 2\text{ }n\right) \left( 3\text{ }n-1\right) ...\left( \frac{%
n+1}{2}\text{ }\frac{n+1}{2}+1\right) .
\end{equation*}
The fixed-point of $\sigma $ is $1$. Under the isomorphism $\iota :\mathbb{Z}%
_{n}\times _{sd}\mathbb{Z}_{2}\longrightarrow D_{n}$, the generators of $%
D_{n}$ are 
\begin{equation*}
\begin{tabular}{lll}
$\iota \left[ \left( g,g\right) \right] =g=a$ & and & $\iota \left[ \left(
g,g^{n-1}\right) \right] =g\sigma =b,$%
\end{tabular}
\end{equation*}
where 
\begin{equation*}
g\sigma =\left( 1\text{ }n\right) \left( 2\text{ }n-1\right) ...\left( \frac{%
n-1}{2}\text{ }\frac{n+1}{2}+1\right) .
\end{equation*}
The fixed-point of $g\sigma $ is $\frac{n+1}{2}$. The Cayley digraph $%
Cay\left( D_{n},\left\{ a=g,b=g\sigma \right\} \right) $ is the $1$-skeleton
of an $n$-gon prism. Let $\rho _{reg}$ be the (right)\ regular permutation
representation of $\mathbb{Z}_{n}$. By Proposition 1,
\begin{equation*}
M\left( Cay\left( D_{n},\left\{ a,b\right\} \right) \right) =\left[ 
\begin{array}{cc}
\rho _{reg}\left( g\right)  & \rho _{reg}\left( g^{n-1}\right)  \\ 
\rho _{reg}\left( g\right)  & \rho _{reg}\left( g^{n-1}\right) 
\end{array}
\right] .
\end{equation*}

\subsection{Line digraphs and in-split graphs}

The notion of split graph is fundamental in symbolic dynamics (see, \emph{%
e.g.}, \cite{LM95}). Let $D$ be a digraph. Then $\overrightarrow{L}D$ is is
a special case of \emph{split graph} of $D$, namely a \emph{in-split graphs}%
. Here is the definition. For every $v_{i}\in V\left( D\right) $, let 
\begin{equation*}
N^{-}\left( v_{i}\right) =I\left( 1,v_{i}\right) \uplus I\left(
2,v_{i}\right) \uplus \cdots \uplus I\left( m\left( v_{i}\right)
,v_{i}\right) ,
\end{equation*}
where $m\left( v_{i}\right) $ is the number of classes in the partition of $%
N^{-}\left( v_{i}\right) $. Let $\mathcal{P}$ be a partition of $A\left(
D\right) $ as above. The \emph{in-split graph} \emph{formed} from $D$ \emph{%
using} $\mathcal{P}$ is denoted by $D_{\left[ \mathcal{P}\right] }$ and
defined as follows: 
\begin{equation*}
V\left( D_{\left[ \mathcal{P}\right] }\right) =\left\{ I\left(
j,v_{i}\right) :v_{i}\in V\left( S\right) ,1\leq j\leq m\left( v_{i}\right)
\right\} ;
\end{equation*}
the number of arcs from the vertex $I\left( k,v_{l}\right) $ to the vertex $%
I\left( j,v_{i}\right) $ is the number of arcs in $D$ which belong to $%
I\left( j,v_{i}\right) $ and have $v_{l}$ as tail. If $\mathcal{P}$ is the
maximal partition (all its classes have cardinality 1) then 
\begin{equation*}
D_{\left[ \mathcal{P}\right] }=\overrightarrow{L}D.
\end{equation*}

\textbf{Acknowledgements.} I wish to thank Yaokun Wu for referring me to 
\cite{LM95}. I wish to thank Richard Jozsa and Andreas Winter for reading
earlier versions of the paper.

\end{document}